\documentclass{moriond}

\def\be{\begin{equation}}
\def\ee{\end{equation}}
\def\bea{\begin{eqnarray}}
\def\eea{\end{eqnarray}}

\usepackage{bm}

\newcommand{\Photo}{\includegraphics[height=35mm]{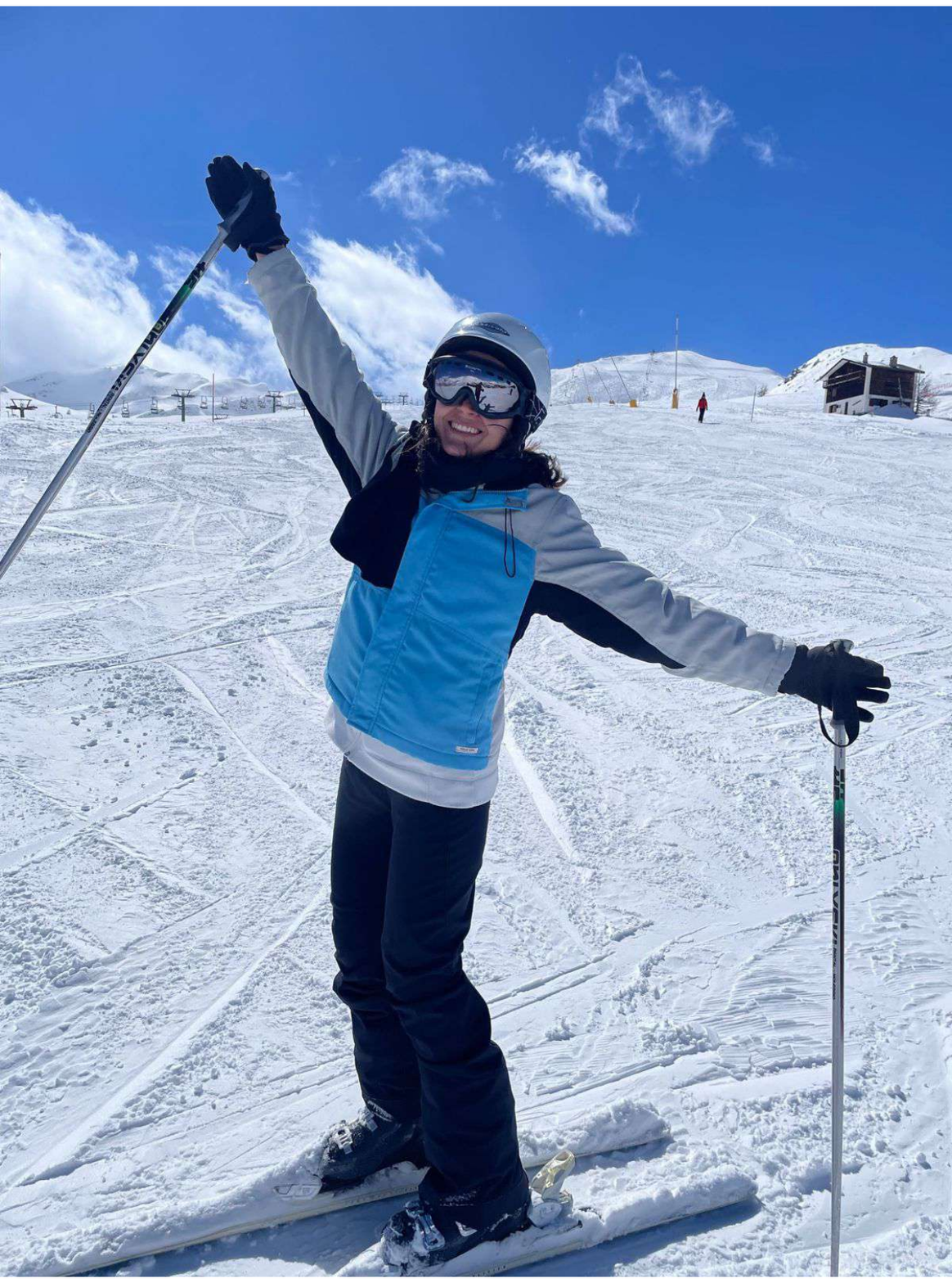}}

\begin{document}
\vspace*{4cm}
\title{Cosmological tests with bright and dark standard sirens}

\author{ ISABELA S. MATOS \footnote{isabela.matos@ictp-saifr.org}}

\address{Instituto de Física Teórica, Universidade Estadual Paulista \& ICTP South American Institute for Fundamental Research, São Paulo 01140-070, SP, Brazil.}

\maketitle\abstracts{
Gravitational waves (GWs) are signals that propagate across large distances in the Universe, and thus, they bring information on the cosmic history. GW sources are at the same time distance indicators and tracers of the matter field. Events generated by binary systems can be divided into bright standard sirens, when followed by electromagnetic transients from which the redshift of the source can be measured, and the more numerous dark standard sirens, when counterparts are not available. In this proceeding, I will discuss some methods for testing the cosmological model using either bright or dark sirens and their combinations with other cosmological probes, focusing on some of my own recent contributions.}

\section{Introduction}

\noindent

Current gravitational wave (GW) detectors LIGO/Virgo have now seen more than a hundred resolved GW sources, from which the majority is identified as coalescing binary black holes (BBHs), while only one event had multi-messenger follow-ups detected in several wavelengths. This channel of observation is very important for cosmology essentially since it allows distance measurements up to high redshifts and without the need of calibration with the cosmic distance ladder. The tricky side of `GW cosmology' comes, however, from the fact that is not possible in general to infer the redshifts of the sources only from the GW signal. That makes the multi-messenger events, the so-called `bright standard sirens' so special, since redshifts can be directly obtained from the electromagnetic counterparts. The current bad ratio of one in a hundred (or two) for bright over dark sirens has urged the community in the last few years to develop techniques for cosmological inference from the much more numerous dark sirens. Among the most successful methods, one relies on the combination with galaxy catalogs, where the information of the large-scale structure of the Universe enters as a sort of `prior' in the redshift distribution of the coalescing binaries \cite{Gair:2022zsa}.

An interesting property of standard sirens is that, as compared to standard candles or rulers, the `GW distance' contains information on the underlying theory of gravity beyond the integrated expansion history $\int_0^z dz'/H(z')$. In most theories of gravity, including all Horndeski family and others, GWs experience a modified friction on large scales that in general enhances (or dim) its amplitude at the observer, and thus, changes the apparent distance to the source $D_{\rm GW}$ as compared to the luminosity distance $D_L$ at the same redshift in the same cosmology\cite{Belgacem2018}:
\begin{equation}
	\Xi(z) := \frac{D_{\mathrm{GW}}(z)}{D_L(z)} \neq 1\,. 
	 \label{gw_distance}
\end{equation}
That means, whatever method is employed, reconstructions of the Hubble diagram with standard sirens (i.e. of $D_{\rm GW}(z)$) are sensitive to, first, the background parameters that are already in the standard $\Lambda$CDM model, $H_0, \Omega_m$ (specially $H_0$), to a possibly evolving dark energy (DE) equation of state $w_{\rm DE}$, and to this unknown function $\Xi(z)$, that is one of the few ingredients for completely specifying most of DE models proposed so far \cite{Bellini2014}. Moreover, going beyond Hubble diagram, specially when combined with other cosmological probes, standard sirens allow testing fundamental properties, like how transparent the cosmos is, and can also be explored as another tracer of the matter distribution. 





\section{Bright standard sirens}

\noindent




Various studies have now forecasted the constraining power of bright siren detections with third generation observatories on measuring $H_0$ and deviations in the GW propagation. We can expect a percent-level measurement of $H_0$ (or better) from bright sirens within a few years of the Einstein Telescope alone. The GW friction could be detected also with percent-level precision if the cosmological background parameters $H_0$ and $\Omega_m$ can be fixed by other observations \cite{Matos2021} \cite{Matos2023}. However, when allowing a generic evolution for $\Xi$ and free cosmological parameters, as argued in \cite{Matos2023}, we do not expect to be able to confidently detect deviations from the standard model only from the Hubble diagram built with bright sirens due to intrinsic strong degeneracies.

This is why in \cite{Matos:2023jkn} we instead used various future distance measurements to check if we can directly measure $\Xi$ independently of the evolution of the background cosmology and with minimal assumptions. Future surveys promise precise measurements of luminosity and angular diameter distances, from upcoming observations of supernovae and the large-scale structure. Together with the brigh sirens, two independent ratios can be probed:
\begin{equation}
	\eta := \frac{D_{L}}{D_A(1+z)^2}\,, \quad \zeta := \frac{D_{\rm GW}}{D_A(1+z)^2}\,.
\end{equation}
The first ratio is the cosmic opacity parameter, which measures violations of the distance duality relation if $\neq 1$ via, e.g., losing photons along the line of sight. Even if defined in terms of angular diameter distance, the second ratio is precisely tracking any deviations in the GW friction ($\Xi$). We do not assume any particular parametrization for the time evolution of $\eta$ or $\zeta$, but assume that they are unity at sufficiently small $z$, so we can `calibrate' the distances with each other.

\begin{figure}
	\centering
	\includegraphics[width=0.98\linewidth]{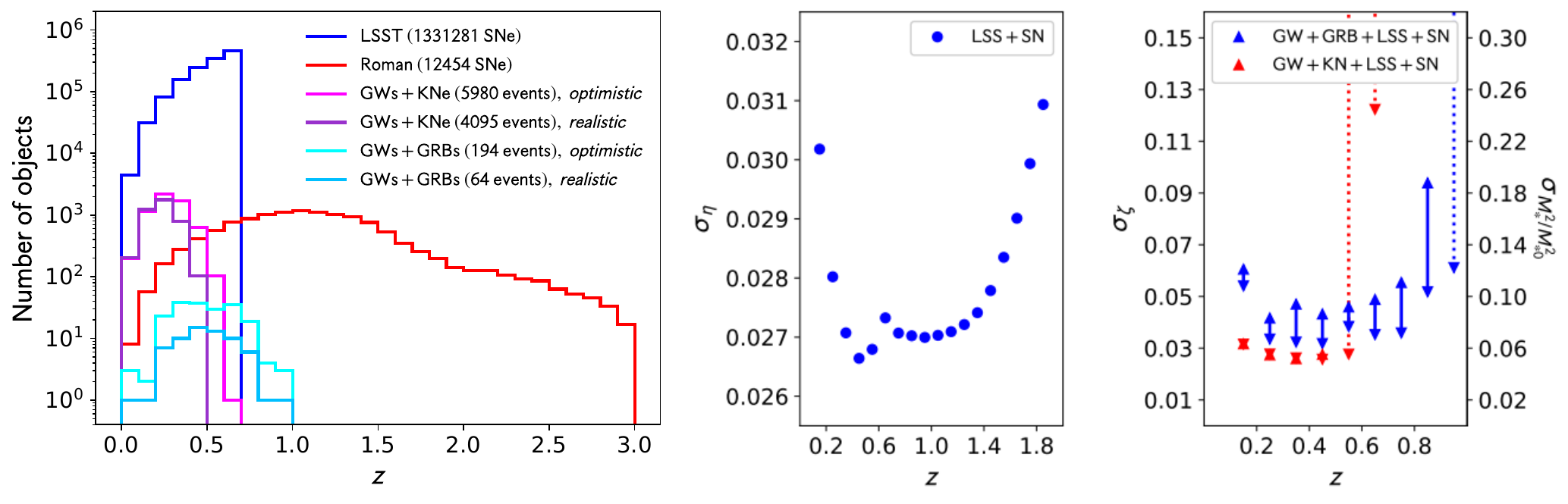}
	\caption{Left: Distribution of events expected for future surveys from which we can get either luminosity or GW distances, and redshifts. Right: Forecasted relative errors on the cosmic opacity parameter $\eta$ and the DE parameter $\zeta$ for each redshift bin, and the corresponding errors in the Planck mass in Horndeski. Figures from ~\protect\cite{Matos:2023jkn}.}
	\label{fig:tripartite}
\end{figure}

We used forecasts for bright sirens with the Einstein Telescope, either expecting a kilonova counterpart detectable with LSST or a short gamma-ray burst (GRB) with the proposed mission THESEUS. As shown in fig. \ref{fig:tripartite}, GRB counterparts as compared to KNe will likely be seen up to higher redshifts due to their intensity, but less often due to the need of having the collimated jet point almost towards us. For luminosity distances we forecasted observations of supernovae with LSST and Roman, from which one can measure $D_L$ up to a constant. This could be fixed with calibration with the distance ladder, which we avoid. Finally, we have also forecasted 5 years of a joint Euclid+DESI-like survey, from which we will able to get angular diameter distance measurements using the model-independent approach called `FreePower' \cite{Amendola:2023awr}. This consists in measuring distances (and also $H(z)$) up to a constant reference value solely from the effects of Alcock-Pazczynski and redshift space distortions in the power spectrum and the bispectrum, by measuring various multipoles, without assuming a particular power spectrum shape, and thus independent of the early Universe physics (or the sound horizon).

Our main results are summarized in the right panel of fig. \ref{fig:tripartite}. We concluded that upcoming measurements of cosmic distances will give us simultaneous percent-level detections of both modified GW propagation and cosmic opacity at various redshifts, independently of the cosmological model and on the specific time evolution of each of these two effects.

\section{Dark standard sirens}

\noindent


When direct redshift measurements are not available, which is the case of the dark standard sirens, indirect methods can be employed. For binaries with at least one neutron star, one possibility is to measure tidal deformabilities that break the degeneracy between mass and redshift in the waveform, the so-called `love sirens'. Another method is to use `known' features in the mass distribution of compact objects, such as gaps, to get redshift information through the redshifted chirp mass, what is called `spectral sirens'. The  so far most explored methods rely, however, on combinations with galaxy catalogs, building on the basic fact that coalescing binaries live in galaxies. They are summarized below.

\begin{itemize}
	\item []{\bf Galaxy catalog or `prior' method.} Here (see, e.g., \cite{Gair:2022zsa}) the redshift of a GW source is estimated by taking all potential host galaxies in a given catalog that lie in the sky region of higher probability for the binary's position, and weighting their redshifts with the probability of each galaxy being the true host, a function that has to be somehow modelled. This `prior' information on the redshift combined with the distance measurement gives a multimodal wide distribution for $H_0$, that becomes informative, however, when several events are put together. 
	In fact, such dark siren method can reach similar precision in $H_0$ than the single bright siren from 2017 depending on the analysis.
	
	There are a few caveats in this method. A first issue is the large sky area inferred from current GW detectors, implying many galaxies have to be accounted for and the computational cost is high -- that will improve with a network of 3g  detectors. Also, current galaxy catalogs are incomplete at significant redshifts and that has to be carefully modelled in order to avoid biases. Furthermore, for those binaries whose true host is not in the catalog, one has to give a prior on the redshift based on the knowledge of the compact objects population model via the merger rate. That is an unknown function introducing more parameters that have to be marginalized over \cite{Gray:2023wgj}.

	\item []{\bf Cross-correlation methods.} Here one computes the cross-correlations between the galaxy and the BBH distributions to reconstruct the Hubble diagram, and possibly also measure the growth of structures. In the method suggested in \cite{Oguri2016} and other works, one can split all GW sources that are observed in distance space in shells of distance, while doing the same for galaxies in the redshift space, and compute their correlations among two arbitrary shells $(z_i, D_{\rm GW, j})$, building the matrices $C_{\ell}(z_i, D_{\rm GW, j})$. When we happen to take shells that indeed correspond to objects at the same distance, that is, $D_{\rm GW, j} = D^{\rm true}_{\rm GW}(z_i)$ in the true underlying cosmology, we should see a peak in the cross-correlation angular power spectrum coefficients. The position of this peak gives, therefore, the Hubble diagram.
	
\end{itemize}

In the ongoing project \cite{Ferri2023prep}, we are prospecting how precisely one can reconstruct the Hubble diagram with future GW data, only from the position of the peak extracted from the data, in a model-independent way. That should bypass several of the caveats from the `prior' method, since we do not expect incompleteness of galaxy catalogs to create biases and no knowledge of population properties seems to be important.

\begin{figure}
	\centering
	\includegraphics[width=0.4\linewidth]{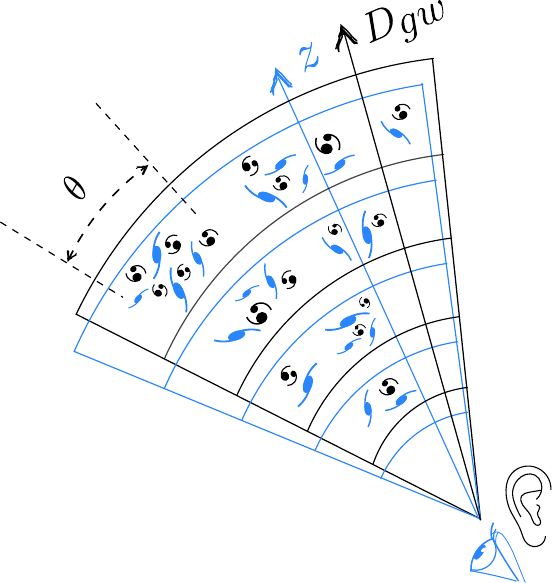}
	\caption{Illustration of the coincident angular clustering of BBHs ad galaxies, represented in distance and redshift spaces, respectively.}
\end{figure}



\noindent

\section*{Acknowledgements}

I would like to thank Rencontres de Moriond for the conference grant and Fundação de Amparo à Pesquisa do Estado de São Paulo (FAPESP) for the PostDoc fellowship nº 2023/02330-0.


\footnotesize

\section*{References}

\bibliography{my_bib}








\end{document}